# Cylindrical Cloak with Axial Permittivity/Permeability Spatially Invariant

Yu Luo, Jingjing Zhang , Hongsheng Chen *, Sheng Xi, and Bae-Ian Wu

*The Electromagnetics Academy at Zhejiang University, Zhejiang University, Hangzhou 310058, P. R. China, and Research Laboratory of Electronics, Massachusetts Institute of Technology, Cambridge, Massachusetts 02139*

**Abstract**

In order to reduce the difficulties in the experimental realizations of the cloak but still keep good performance of invisibility, we proposed a perfect cylindrical invisibility cloak with spatially invariant axial material parameters. The advantage of this kind of TE (or TM) cloak is that only $\rho$ and $\varphi$ components of $\mu$ (or $\varepsilon$) are spatially variant, which makes it possible to realize perfect invisibility with two-dimensional (2D) magnetic (or electric) metamaterials. The effects of perturbations of the parameters on the performance of this cloak are quantitatively analyzed by scattering theory. Our work provides a simple and feasible solution to the experimental realization of cloaks with ideal parameters.

*Author to whom correspondence should be addressed; electronic mail: chenhs@ewt.mit.edu

Based on form-invariant coordinate transformations, cloak of invisibility was proposed which can perfectly conceal arbitrary objects from the detection [1,2]. Similar conformal mapping method was applied to 2D space to produce similar effects in geometric limit [3]. Following this approach, the experimental realization of cylindrical cloak at microwave frequency was soon reported [4], and increasing interest on cloak has been raised afterwards [5-20] One problem in cloak design is that the ideal cylindrical cloak proposed in Ref [5] has three spatial variant material parameters, which is very difficult to realize in practice. Cylindrical cloak designed with simplified parameters are proposed aiming to decrease the complexity in the realization of cloak [4,5,9]. However, Theoretical [14] and numerical [16,17] results show that the scattering of such kind of cloaks can not be totally removed. In order to reduce the difficulties in the experimental realizations of the cloak but still keep good performance of invisibility, in this letter, we provide a feasible solution to realize a kind of perfect cylindrical cloak where the axial ($z$ direction) components of the material parameters are larger than 1 and spatially invariant. The advantage of this kind cloak is that only $\rho$ and $\varphi$ components of the relative $\mu$ (or $\varepsilon$) are space dependent, which makes it possible to realize perfect (or near perfect) invisibility with two-dimensional (2D) magnetic (or electric) metamaterials. We analytically demonstrated the perfect invisibility of this kind of cloak as well as the sensitivity to perturbations. With the coming forth of more and more 2D metamaterial structures [21-23], cloaks with the proposed ideal parameters are hence brought realizable.

We use TE cylindrical cloak as an example and start our discussion by considering a scale transformation in cylindrical coordinate (TM case can be similarly derived): $\rho' = f(\rho)$, $\varphi' = \varphi$, $z' = z$, where $f$ is an arbitrary smooth function. Under the above mapping, the parameters of the transformation media can be expressed as

$$\mu_\rho = \frac{f(\rho)}{\rho f'(\rho)}, \quad \mu_\varphi = \frac{\rho f'(\rho)}{f(\rho)}, \quad \varepsilon_z = \frac{f'(\rho) f(\rho)}{\rho}. \tag{1}$$

In order to make the axial component spatially invariant, we let $\varepsilon_z = \frac{f'(\rho) f(\rho)}{\rho} = C$, where $C$ is an undetermined coefficient. Since the transformation of perfect cloak requires $f(R_1) = 0$ and $f(R_2) = R_2$, the solution for $f(\rho)$ can be obtained. We insert it into Eq. (1) and get the following material parameters of the cloak

$$\mu_\rho = \frac{\rho^2 - R_1^2}{\rho^2}, \quad \mu_\varphi = \frac{\rho^2}{\rho^2 - R_1^2}, \quad \varepsilon_z = \frac{R_2^2}{R_2^2 - R_1^2}. \tag{2}$$

It is interesting to see that the relative permittivity $\varepsilon_z$ is spatial invariant and larger than 1. Since only $\mu_\rho$ and $\mu_\varphi$ are spatially variant, this cloak can be realized with 2D metamaterials like SRRs of honeycomb [21] or cross-embedded arrangements [22,23] but with different size in the $\rho$ and $\varphi$ directions. In theory, as long as the parameters of the metamaterials take the exact ideal forms addressed in equation (2), cloak with perfect invisibility can be realized. However, some small perturbations can still deteriorate the performance of the cloak due to the extreme values of $\mu_\rho$ and $\mu_\varphi$ (zero or infinity) at the interior surface. In addition, the electric response of 2D magnetic ring resonators (for instance, SRRs) will affect the bulk effective permittivity, resulting in minor changes to $\varepsilon_z$. Therefore, it is necessary to make theoretical analysis on how the perturbations quantitatively affect the performance of the cloak.

By including some perturbations to Eqs. (2), we consider a more general case where the parameters of cloak are given by

$$\mu_{\rho} = \frac{a\left(\rho^2 - R_1^2\right)}{\rho^2}, \quad \mu_{\varphi} = \frac{a\rho^2}{\rho^2 - R_1^2}, \quad \varepsilon_z = \frac{R_2^2}{R_2^2 - R_1^2}(1+\sigma) = b, \tag{3}$$

where $a$, $b$ and $\sigma$ are all constants. $\sigma$ represents the perturbation of $\varepsilon_z$ due to the electric resonance of 2D magnetic ring resonators and $a$ is controllable by artificial metamaterials. The inner boudary of the cloak shell is set at $R_{in} = R_1 + \Delta R$, where $\Delta R$ is a small positive number, so that extreme material parameters at the interior surface can be avoided [13]. The cloak consists of a PEC lining at the interior surface. The performance of this cloak can be examined by cylindrical wave expansion method. For TE polarized wave, the general wave equation governing the $E_z$ field in the cloak layer is

$$\frac{1}{\rho}\frac{\partial}{\partial \rho}\frac{\rho^2 - R_1^2}{\rho}\frac{\partial}{\partial \rho}E_z + \frac{1}{\rho^2 - R_1^2}\frac{\partial^2}{\partial \varphi^2}E_z + abk_0^2 E_z = 0. \tag{4}$$

Utilizing separation of variables method, the solution of Eq. (4) can be obtained

$$E_z = B_n\left(k_0 g(\rho)\right)e^{in\varphi}, \tag{5}$$

where $g(\rho) = \sqrt{ab\left(\rho^2 - R_1^2\right)}$, and $B_n$ represents the solutions of Bessel function of the $n$-th order. Suppose a TE polarized plane wave with unit amplitude $E^i = \hat{z}e^{ik_0 x}$ is incident upon the cloak along $x$ direction. The incident fields, scattered fields and the fields inside the cloak layer can be described as

$$E_z^i = \sum_{n=-\infty}^{\infty} a_n J_n(k_0 \rho)\cos n\varphi, \tag{6}$$

$$E_z^s = \sum_{n=-\infty}^{\infty} b_n H_n^{(1)}(k_0 \rho)\cos n\varphi, \tag{7}$$

$$E_z^c = \sum_{n=-\infty}^{\infty} \left[c_n J_n\left(k_0 g(\rho)\right) + d_n N_n\left(k_0 g(\rho)\right)\right]\cos n\varphi. \tag{8}$$

Here $a_n = i^n$. $b_n$, $c_n$ and $d_n$ are unknown expansion coefficients. $J_n$ $N_n$ and

$H_n^{(1)}$ represent the *n*-th order of Bessel function, Neumann function, and Hankel function of the first kind, respectively. By applying boundary condition (continuity of $E_z$ and $H_\varphi$) at the inner and outer interface of the cloak, all the coefficients can be determined

$$b_n = \frac{BJ_n(k_0R_2) - AJ'_n(k_0R_2)}{AH_n^{(1)'}(k_0R_2) - BH_n^{(1)}(k_0R_2)}, \tag{9}$$

$$c_n = \frac{J_n(k_0R_2)H_n^{(1)'}(k_0R_2) - J'_n(k_0R_2)H_n^{(1)}(k_0R_2)}{AH_n^{(1)'}(k_0R_2) - BH_n^{(1)}(k_0R_2)}, \tag{10}$$

$$d_n = -\frac{J_n(k_0g(R_{in}))}{N_n(k_0g(R_{in}))}c_n, \tag{11}$$

where $A = J_n(k_0g(R_2)) - \frac{J_n(k_0g(R_{in}))}{N_n(k_0g(R_{in}))}N_n(k_0g(R_2))$ and

$$B = \sqrt{\frac{b(R_2^2 - R_1^2)}{aR_2^2}}\left(J'_n(k_0g(R_2)) - \frac{J_n(k_0g(R_{in}))}{N_n(k_0g(R_{in}))}N'_n(k_0g(R_2))\right).$$

When $a = 1$, $\sigma = 0$ and $\Delta R = 0$, Eqs (9)-(11) can be simplified to $b_n = d_n = 0$, $c_n = a_n = i^n$. The exactly zero value of the scattering coefficient $b_n$ indicates that the cloak achieved with the parameters given by Eq. (2) is perfect. However, $\mu_\varphi$ is infinite at the inner boundary of the cloak, which is very difficult to achieve in practice. Therefore, a theoretical discussion of the perturbations at the cloak's interior surface, as well as how the loss affects the cloaking performance is essential. Using the asymptotical approach raised in Ref [13], we calculated the total field distributions for both lossless and lossy cases with a perturbation of $\Delta R = 0.1R_1$. Comparing Fig. 1(a) and Fig. 1(b), which correspond to the field distribution for lossless and lossy cloaks respectively, we find that noticeable scattered field is induced by the perturbation when the cloak are lossless, but introducing the loss can suppress the scattering in almost all the direction (expect for a small cone in the forward direction). Fig. 1(c) depicts the normalized differential scattering efficiency [15] for the imperfect cloaks ($\Delta R = 0.1 R_1$) with different loss tangent ($\tan(\delta)$=0, 0.1 and 0.5) introduced to $\mu_\rho$ and $\mu_\varphi$, showing that the backward scattered power decreases as the loss increases. In order to further look into this problem, we also calculate the total scattering cross sections (the integral of scattering power from 0° to 360°) as a function of loss tangent for three imperfect cloaks with different perturbation at the interior surface ($\Delta R = 0.1 R_1, 0.2 R_1$, and $0.4 R_1$), as shown in Fig. 2. It is interesting that for every definite $\Delta R$, we can find a loss tangent which can minimize the total scattering cross section (for instance, when $\Delta R = 0.1 R_1$, introducing a loss tangent $\tan(\delta) = 0.08$ can minimize the total scattering cross section). The reason is that when small loss is introduced, the field that penetrates into the inner boundary of the cloak decreases, therefore, the effect of perturbation in the inner boundary also decreases. Our theoretical analysis on the cloaks with other linear transformations [5, 9, 13] also show the similar property in that properly introducing loss to the parameters can improve the performance of the asymptotically linear cloak, making it insensitive to the perturbation at the interior surface.

We next study how sensitive the cloaks are to the material parameters. We calculate the total field distributions of imperfect cloaks when $\varepsilon_z$ slightly changes from ideal parameters, i.e. $\sigma = 0.2$, under three different conditions. Fig. 2(a) depicts the $E_z$ field distribution with ideal $\mu_\rho$ and $\mu_\varphi$ ($a=1$); Fig. 2(b) is the case where the impedance $\sqrt{\dfrac{\mu_\varphi}{\varepsilon_z}}$ and $\sqrt{\dfrac{\mu_\rho}{\varepsilon_z}}$ are the same to those of the perfect cloak

($a=1+\sigma$). And in Fig. 2(c) the refractive index $\sqrt{\mu_\varphi \varepsilon_z}$ and $\sqrt{\mu_\rho \varepsilon_z}$ are kept unchanged ($a=\frac{1}{1+\sigma}$) everywhere. The normalized differential cross sections for the above three imperfect cloaks are shown in Fig. 2 (d). The blue dot-dashed line, the black dashed line and the red solid line correspond to the cloaks with ideal permeability, ideal impedance and ideal refractive index, respectively. It is obvious that when there is small perturbation to $\varepsilon_z$, keeping the refractive index the same as the ideal one can suppress the scattering ominidirectionally, while letting the impedance of cloak remain ideal yields an even better performance in backward scattering. Therefore, for bistatic (transmitter and receiver in different locations) detection the imperfect cloak should be achieved with ideal refractive index, and for monostatic detection (transmitter and receiver in the same locations) the imperfect cloak should be realized with the same impedance as the perfect one [4, 5, 13].

In conclusion, we have proposed a perfect cylindrical cloak for TE (TM) wave with only two spatially variant parameters. The scattering properties of this cloak with a perturbation at the inner boundary are studied with cylindrical wave expansion method, indicating that both backward scattering and total scattering cross section of this imperfect cloak can be suppressed by appropriately introducing loss. It is also found that when a tiny change of $\varepsilon_z$ is introduced but the refractive index remains the same, the cloak can still reduce the scattering cross section omnidirectionally, while keeping the impedance unchanged can yield an even better performance in concealing objects from monostatic detection. With the advent of more and more 2-D metamaterial structures, the approaches we proposed here can be easily implemented.

This work is sponsored by the Chinese National Science Foundation under Grant Nos. 60531020, 60671003 and 60701007, the NCET-07-0750, the ZJNSF (R105253),

## Figure Captions

FIG. 1 (color online) (a) $E_z$ field distribution due to a tiny perturbation at the interior surface of the cloak $R_{in} = R_1 + 0.1R_1$ when the parameters are lossless. (b) The same field distribution for the imperfect cloak when the loss tangent of $\mu_\rho$ and $\mu_\varphi$ are 0.5 (the red dashed line in (a) and (b) indicate the position of perfect interior surface $R_{in} = R_1$). (c) The normalized differential cross sections of the imperfect cloaks with a specified loss tangent introduced to $\mu_\rho$ and $\mu_\varphi$.

FIG. 2 (color online) The normalized total scattering cross sections as a function of loss tangent for three imperfect cloaks with different perturbations at the inner boundary.

FIG. 3 (color online) $E_z$ field distribution due to a TE polarized plane wave incident onto the three kinds of near-ideal cloaks when there is a tiny change of $\varepsilon_z$ ($\delta = 0.2$). (a) $\mu_\rho$ and $\mu_\varphi$ are still ideal ($a = 1$). (b) Impedance $\sqrt{\frac{\mu_\varphi}{\varepsilon_z}}$ and $\sqrt{\frac{\mu_\rho}{\varepsilon_z}}$ are the same as perfect cloak ($a = 1+\sigma$). (c) Refractive index $\sqrt{\mu_\varphi \varepsilon_z}$ and $\sqrt{\mu_\rho \varepsilon_z}$ remains unchanged ($a = \frac{1}{1+\sigma}$). (d) The corresponding differential cross sections of these three kinds of imperfect cloaks.

Fig.1.

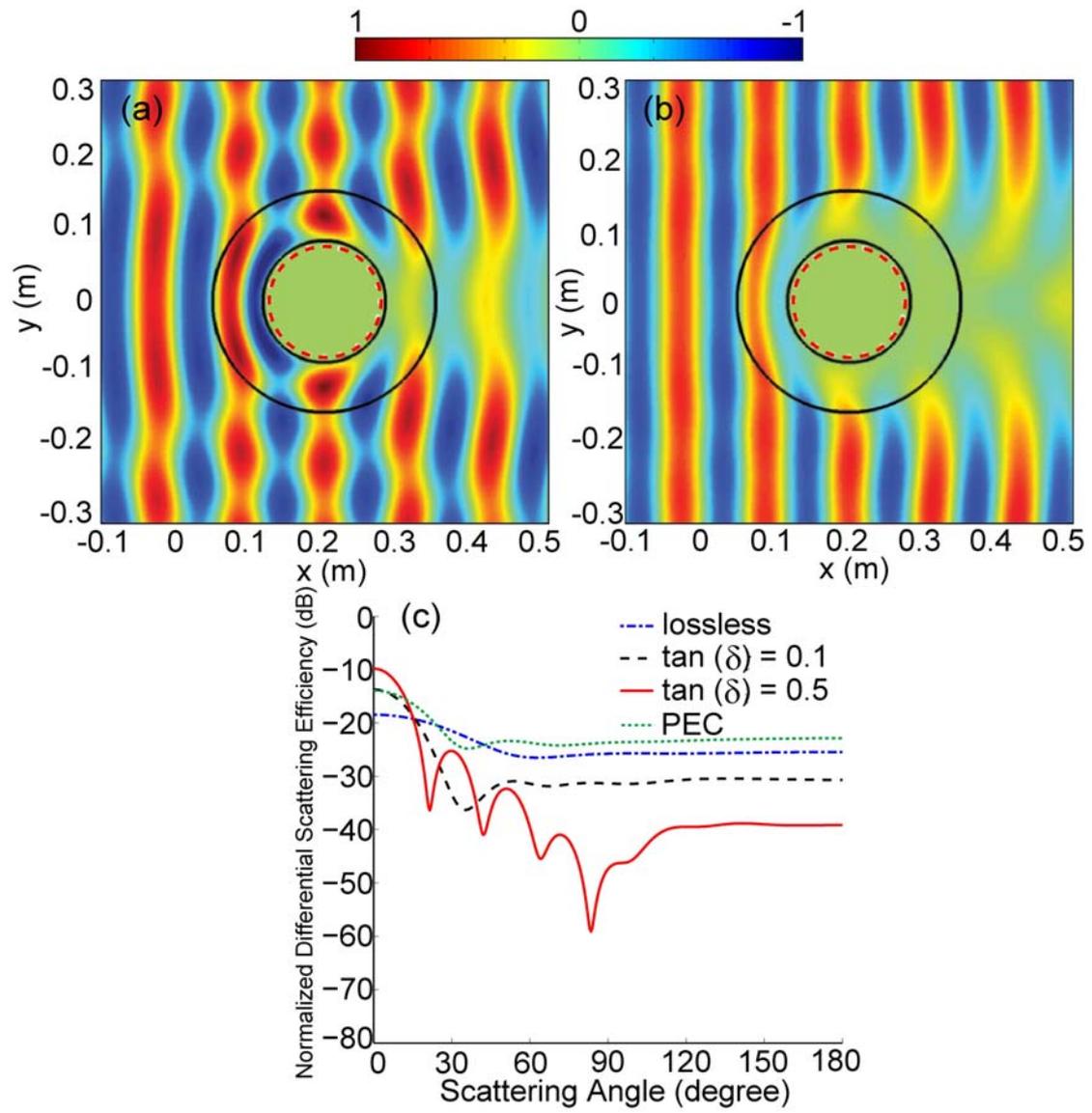

Fig.2

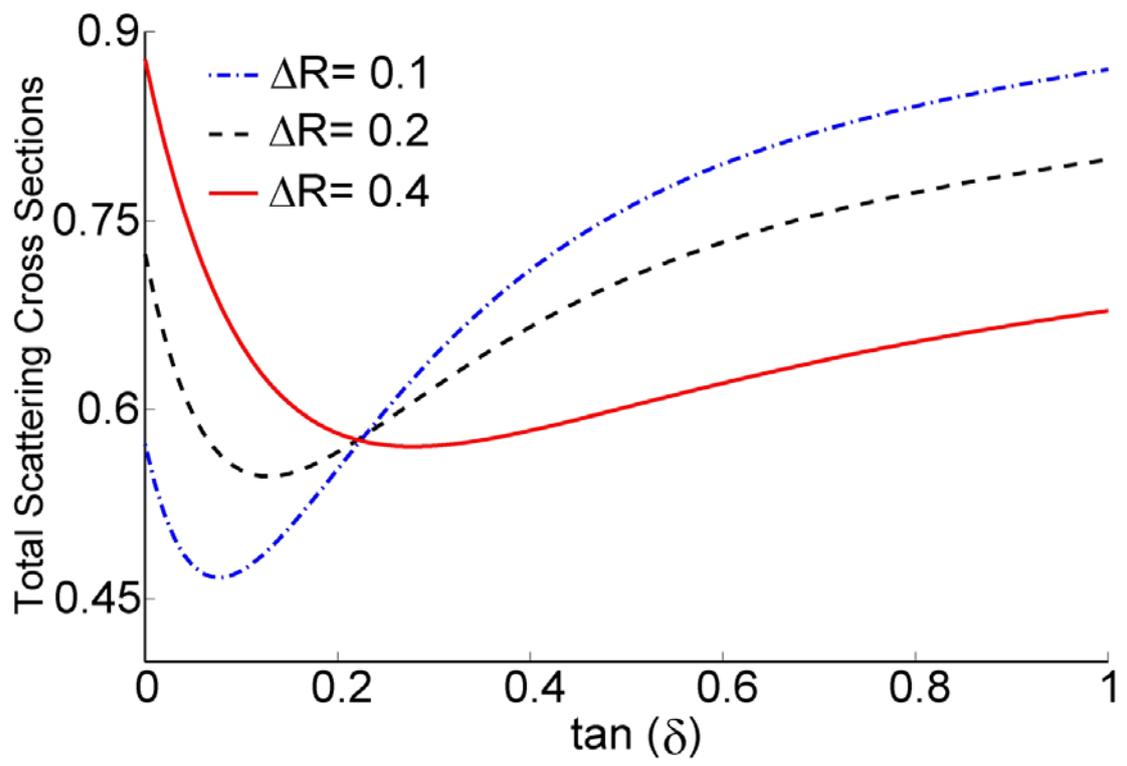

Fig.3

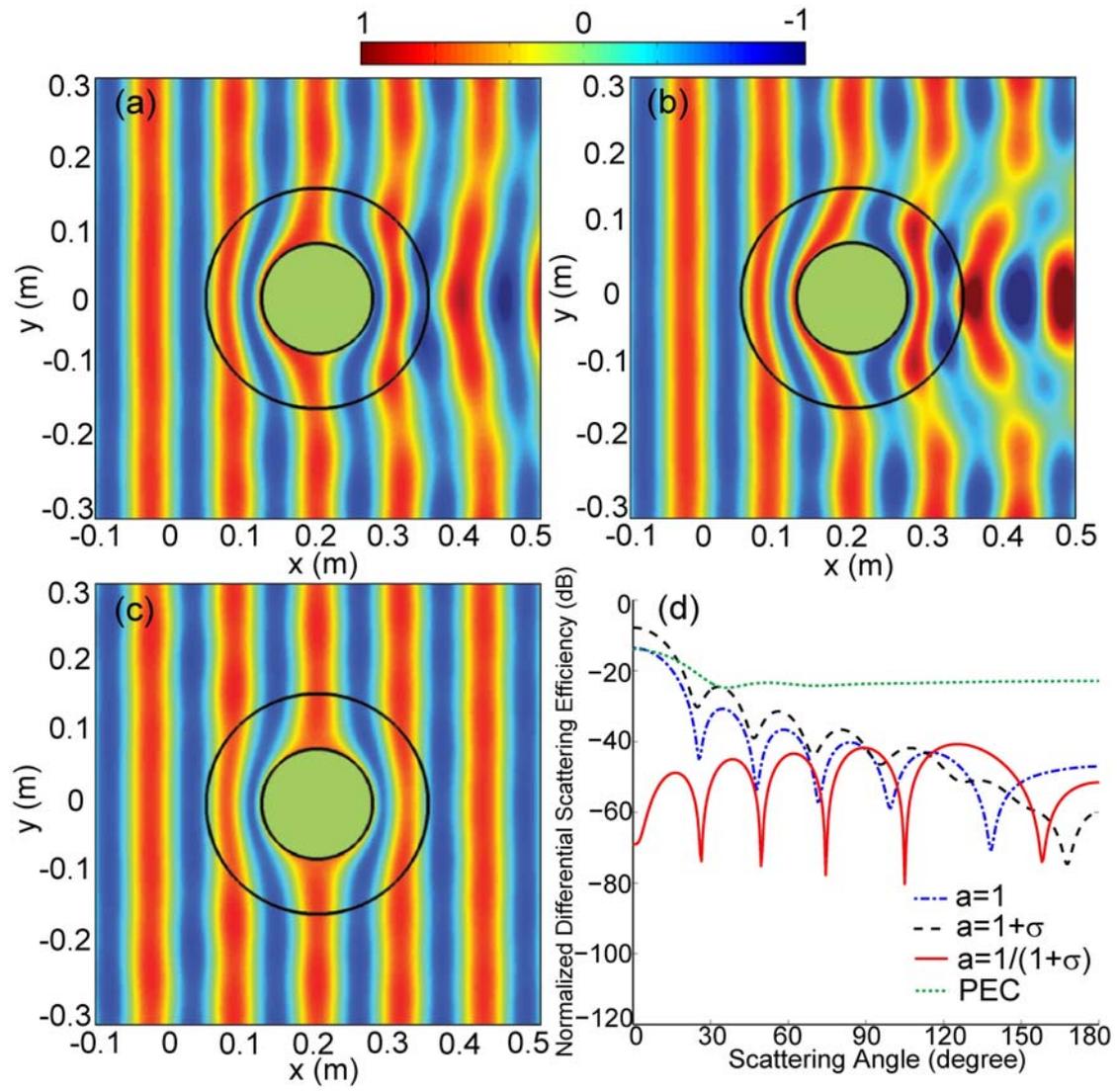